# Universal coherence protection in a solid-state spin qubit

Kevin C. Miao[1], Joseph P. Blanton[1,2], Christopher P. Anderson[1,2], Alexandre Bourassa[1], Alexander L. Crook[1,2], Gary Wolfowicz[1,4], Hiroshi Abe[3], Takeshi Ohshima[3], David D. Awschalom[1,2,4,*]

[1]Pritzker School of Molecular Engineering, University of Chicago, Chicago, IL 60637, USA.

[2]Department of Physics, University of Chicago, Chicago, IL 60637, USA.

[3]National Institutes for Quantum and Radiological Science and Technology, 1233 Watanuki, Takasaki, Gunma 370-1292, Japan.

[4]Center for Molecular Engineering and Materials Science Division, Argonne National Laboratory, Lemont, IL 60439, USA.

*email: awsch@uchicago.edu

**Decoherence largely limits the physical realization of qubits and its mitigation is critical to quantum science. Here, we construct a robust qubit embedded in a decoherence-protected subspace, obtained by hybridizing an applied microwave drive with the ground-state electron spin of a silicon carbide divacancy defect. The qubit is protected from magnetic, electric, and temperature fluctuations, which account for nearly all relevant decoherence channels in the solid state. This culminates in an increase of the qubit's inhomogeneous dephasing time by over four orders of magnitude (to >22 milliseconds), while its Hahn-echo coherence time approaches 64 milliseconds. Requiring few key platform-independent components, this result suggests that substantial coherence improvements can be achieved in a wide selection of quantum architectures.**

Electron spins embedded in a solid-state host, such as silicon carbide (SiC) and diamond, are attractive platforms for quantum information processing by virtue of their optical interface (*1, 2*) and engineered interactions with the host crystal (*3–7*). These electron spins are highly controllable using external magnetic fields because of the electron's large magnetic moment. However, unwanted magnetic field noise will also couple intensely through this degree of freedom and quench spin coherences (*8, 9*). In particular, a naturally abundant host crystal typically has non-zero nuclear spin isotopes and paramagnetic impurities, whose fluctuations produce variations in the magnetic field at the electron spin with magnitudes as large as 0.45 mT (*10*). These magnetic field fluctuations limit the electron spin's Hahn-echo coherence time ($T_2$) to around a millisecond (*7, 8, 11, 12*) and suppress spin inhomogeneous dephasing times ($T_2^*$) to a few microseconds (*11, 12*). These relatively short inhomogeneous dephasing times typically bound the timescales for quantum information storage and manipulation, requiring the development of nuclear spin quantum memories (*13–15*), while also limiting the fidelity of quantum information transfer to coupled systems, such as superconducting resonators (*16, 17*). These restrictions call for the extension of electron spin coherence, which is typically achieved by means of dynamical decoupling (*18, 19*), isotopic purification (*14, 20–22*), or Hamiltonian engineering (*3, 4, 23–27*).

One particular solid-state electron spin system, the basally oriented *kh* divacancy in the 4H polytype of SiC, has recently been shown to have reduced sensitivity to magnetic noise when operated near zero external magnetic field (*28*). This robustness primarily arises from the transverse zero-field splitting (ZFS) intrinsic to the spin system's single mirror plane symmetry (Fig. 1A, lower left). When combined with the spin system's axial ZFS, clock transitions emerge near **B** = 0. By operating at this point, the spin has lengthened inhomogeneous dephasing times, among other properties (see Supplementary Materials). In this work, we implement a protocol to further extend the ground-state spin inhomogeneous dephasing time, as well as the Hahn-echo coherence time, of a single *kh* divacancy in 4H-SiC by simply applying a continuous microwave drive resonant with two of the three ground-state spin-1 sublevels at zero field. This microwave drive induces spin-photon hybridization, producing dressed spin states (*3*, *4*, *23–27*) that exhibit notably decreased inhomogeneous shifts from nearly all environmental fluctuations. This leads to exceptionally long spin coherence times in a decoherence-protected subspace (DPS). While many dressing protocols decouple the target system from a subset of noise sources, we show that the dressed spin states presented here, by virtue of the system's large transverse and axial ZFS magnitudes, are protected from magnetic, electric, and temperature fluctuations to high order. Furthermore, there are many practical advantages of this protocol when compared to typical pulsed dynamical decoupling sequences, including uninterrupted coherence protection, lower peak drive powers, and fewer non-idealities in implementing the protocol. Importantly, this dressed spin system retains its large magnetic and electric response to resonant microwave control fields, preserving the capability to rapidly perform quantum operations to coherently manipulate the long-lived quantum state and the ability to prepare arbitrary superpositions in this new basis.

We isolate the *kh* divacancy used in this study in naturally abundant, commercially available 4H-SiC at 5 K (see Methods). The ground-state spin-1 system is initialized and read out through optical pulses, while spin rotations are produced by microwave-frequency magnetic and electric fields from on-chip wires and capacitors, respectively (Fig. 1A). We use a three-axis electromagnet for vector control of the external magnetic field along the Cartesian axes of the spin system (Fig. 1A, lower left). At **B** = 0, a spin transition can be driven between the upper two spin states, $|+\rangle = \frac{1}{\sqrt{2}}(|+1_z\rangle + |-1_z\rangle)$ and $|-\rangle = \frac{1}{\sqrt{2}}(|+1_z\rangle - |-1_z\rangle)$ (where $|\pm 1_z\rangle$ denote the $m_s = \pm 1$ sublevels in the $S_z$ basis). We use Ramsey interferometry of the $|+\rangle \leftrightarrow |-\rangle$ transition to determine the transverse ZFS magnitude $E/(2\pi)$ = 18.353164(4) MHz to a high degree of precision and accuracy (see Supplementary Materials), allowing us to apply a continuous microwave-frequency magnetic drive at an angular frequency $\omega = 2E$ resonant with the $|+\rangle \leftrightarrow |-\rangle$ transition (Fig. 1B). At a sufficiently high drive Rabi frequency $\Omega$, the $|\pm\rangle$ spin levels undergo Autler-Townes splitting (see Supplementary Materials) to form dressed spin states $|\pm 1\rangle$ with energy levels offset from that of $|+\rangle$ by $\pm\Omega/2$. These dressed states form the basis for the DPS, and can be observed by applying a weaker microwave probe pulse and sweeping its frequency detuning $\Delta$ from the $|0\rangle \leftrightarrow |+\rangle$ transition frequency (Fig. 1C). For subsequent measurements, we operate at $\Omega/(2\pi)$ = 350 kHz to mitigate higher-order energy dispersion components (see Supplementary Materials). We realize coherent control within the dressed spin-1 system by driving $\Delta m_s = \pm 1$ transitions $|0\rangle \leftrightarrow |\pm 1\rangle$ with ac magnetic fields (Fig. 1D) and a $\Delta m_s = \pm 2$ transition $|+1\rangle \leftrightarrow |-1\rangle$ with ac electric fields (*29*) (Fig. 1E). Readout of the $|\pm 1\rangle$ basis is accomplished by non-adiabatically disabling the dressing drive, rotating into the $\{|0\rangle, |+\rangle\}$ basis, and optically probing the spin population in $|0\rangle$ (see Supplementary Materials).

We quantify the energy inhomogeneity of this driven spin system by preparing a superposition $|\psi_0\rangle = \frac{1}{\sqrt{2}}(|+1\rangle + |-1\rangle)$ in the DPS and executing a Ramsey free precession sequence. Remarkably, with active feedback procedures in place, we measure a $T_2^*$ spin inhomogeneous dephasing time in this basis to be 22.4(10) milliseconds (Fig. 2A), which is over four and two orders of magnitude longer than the $T_2^*$ measured for the same $kh$ divacancy at $B_z$ = 1.2 mT and $\mathbf{B} = 0$, respectively (Fig. 2B). Moreover, by adding a single refocusing pulse, we extend the lifetime of the superposition to a $T_2$ Hahn-echo spin coherence time of 64(4) milliseconds (Fig. 2C). Both of these values are among the longest times measured for an optically addressable electron spin, regardless of isotopic purity of the host. These increases directly result from the reduced energy inhomogeneity of the spin in the DPS, combined with active feedback of the spin resonance frequency. We discuss these two components in detail in the following sections.

We examine the energy dispersion relations between the dressed spin levels under magnetic, electric, and temperature fluctuations in order to reveal their contributions to inhomogeneity in the DPS. We first consider the effect of magnetic fields on the dressed spin levels, as magnetic noise typically limits the coherence of many electron spin systems. Under a continuous drive resonant with the $|+\rangle \leftrightarrow |-\rangle$ transition, the dispersion relation of the ground-state spin acquires both quadratic and quartic dependences on magnetic field (see Supplementary Materials). We probe the dispersion relation by applying magnetic fields along the $x$- and $z$-axes of the spin system while scanning the frequency detuning $\Delta$ of a probe drive (Fig. 3A,B). Rotational symmetry of the system about the $z$-axis leads to indistinguishable effects from $x$- and $y$-axis fields (see Supplementary Materials). The resulting frequencies of the $|0\rangle \leftrightarrow |\pm 1\rangle$ spin resonances provide us with the driven spin system's spectral response to magnetic fields. To confirm these observations, we develop an analytical model of the driven spin system using Floquet analysis alongside a numerical model through spin-1 master equation simulations (see Supplementary Materials). We find excellent agreement between the experimental data and these models (Fig. 3C,D). We then consider the primary source of magnetic noise, the nuclear spin bath, as an isotropic magnetic fluctuator with an estimated fluctuation magnitude at this $kh$ divacancy to be 13 µT (see Supplementary Materials). We perform fine scans of the $|+1\rangle \leftrightarrow |-1\rangle$ transition using Ramsey interferometry (Fig. 3E,F) and confirm the high degree of insensitivity in this range, illustrating the primary mechanism for the largely increased spin coherence times. These results, once combined with a phenomenological model of the spin energy inhomogeneity (see Supplementary Materials), allow us to quantitatively confirm that residual inhomogeneity from magnetic fields still plays a role in limiting the spin coherence, even after substantial suppression in the DPS.

We can then apply the analytical energy dispersion relations to understand the effect of electric noise and temperature fluctuations on spin inhomogeneous dephasing in the DPS. Electric and temperature fluctuations affect the axial and transverse ZFS magnitudes, as well as a corresponding detuning of the dressing drive from the resonance frequency $\omega = 2E$. We use the undriven spin's first-order sensitivity to electric fields (*30*) at $\mathbf{B} = 0$ to quantify the magnitude of electric field noise present in the system (see Supplementary Materials). In the DPS, we gain first-order protection against fluctuations in the ZFS magnitudes, resulting in increased robustness against electric field noise and temperature shifts. This leads to a reduction of electrically induced spin energy inhomogeneity by nearly two orders of magnitude (see Supplementary Materials), and diminishes electric field contributions in limiting the spin coherence in the DPS. Further reduction of electric field contributions may emerge by applying a dc electric field to deplete fluctuating charges (*7*, *28*), potentially leading to nearly complete elimination of electric field noise.

The energy dispersion relations in the driven basis confirm that the spin's energy levels depend linearly on the dressing drive Rabi frequency. Hence, amplitude drifts of the dressing drive and corresponding fluctuations in the Rabi frequency $\delta\Omega$ introduce first-order inhomogeneity of the dressed spin levels and cause shortening of the measured $T_2^*$. To this end, we implement active feedback of the dressed spin resonance frequency in order to counteract these slow drifts and reduce inhomogeneity (Fig. 4A). By evaluating an error signal derived from Ramsey free precession of the dressed spin (Fig. 4B), we measure and correct drifts on the order of 30 Hz in the spin resonance frequency (see Methods), consistent with the 100 ppm/°C stability of the dressing drive oscillator. In order to validate the effectiveness of our feedback protocol, we measure Ramsey free precession of $|\psi_0\rangle = \frac{1}{\sqrt{2}}(|+1\rangle + |-1\rangle)$ with and without feedback enabled (Fig. 4C). With active feedback enabled, we measure our reported $T_2^*$ of 22.4(10) ms; when disabled, this value shortens to 17.4(10) ms as the dressing drive inhomogeneity suppresses the spin's coherence on the timescale of hours (Fig. 4D). These results suggest that by incorporating this feedback protocol, we have largely mitigated the effects of dressing drive amplitude fluctuations and that noise sources intrinsic to the host crystal remain as the limiting factor for spin coherence in the DPS.

We have presented a considerable increase in inhomogeneous dephasing time by over four orders of magnitude and a thousand-fold increase in Hahn-echo coherence time of a single *kh* divacancy electron spin under the effect of a single continuous microwave dressing drive tone. The resulting extended spin $T_2^*$ enables resonant coupling to weakly interacting quantum systems, paving the way to hybrid quantum systems using single divacancy ground-state spins. Even with modest coupling rates to the spin, unity-cooperativity systems can be realized. Furthermore, the viability of this protocol at zero magnetic field is a substantial advantage, as superconducting systems with low critical fields can be integrated with the spin system. We have shown that, while insensitive to non-resonant magnetic and electric fluctuations, the electron spin is still highly responsive to resonantly applied magnetic and electric control fields. Thus, prospective hybrid quantum systems involving the driven *kh* divacancy spin can leverage any one of magnetic, electric, and acoustic (*3*, *6*) couplings to the spin. Lastly, megahertz-scale tuning of the dressed spin energy levels allows for rapid adjustment of interaction strengths with other quantum systems, enabling the spin to be an efficient quantum bus.

The few requirements for successful implementation of this protocol allow for immediate extension to other quantum systems with similar undriven energy structures. Spin-1 systems with larger transverse ZFS values can expect to see improvements from higher dressing drive Rabi frequencies and greater suppression of inhomogeneity. Candidate systems include the basally oriented *hk* divacancy in 4H-SiC ($E/(2\pi)$ = 82.0 MHz (*31*)) as well as basally oriented nitrogen-vacancy centers in 4H-SiC ($E/(2\pi)$ = 103 MHz (*32*)) and 6H-SiC ($E/(2\pi)$ = 138 MHz (*33*)). Strained nitrogen-vacancy centers in diamond with an appreciable transverse ZFS (*16*) may also stand to benefit from this protocol. Turning to other systems, donor spins in silicon with robust clock transitions (*34, 35*), superconducting qubits operated at a degeneracy point (*36*), and molecular spins (*37, 38*) may be driven to produce enhanced coherence.

Furthermore, our protocol is compatible with additional forms of noise suppression, including isotopic purification, charge depletion, and pulsed dynamical decoupling. By reducing the volume density of fluctuating spins in the bath and thus reducing the average dipolar interaction strength between the electron spin and bath nuclear spins (*14, 20–22*), we expect the magnitude of residual inhomogeneous electron spin energy shifts to decrease proportionally. Dynamic techniques such as spin bath driving (*12*) and spin bath hyperpolarization (*39*) may further reduce the noise contributions from spin bath fluctuations. Charge depletion, which has led to the observation of near-transform-limited optical linewidths in divacancies (*7, 28*), can further suppress electric field contributions. In the present work, we have demonstrated compatibility with pulsed dynamical decoupling using a Hahn-echo sequence; higher-order dynamical decoupling sequences may allow for spin coherence to rapidly reach the spin-lattice relaxation limit with relatively few control pulses. Considering these prospective avenues, this dressing protocol represents a major development for potentially improving coherence across a variety of quantum systems, and is a crucial step towards integrating SiC divacancies into robust, real-world quantum technologies.

**Acknowledgements:** We thank M. Fukami, S. L. Bayliss, Y. Tsaturyan, P. C. Jerger, V. V. Dobrovitski, M. Onizhuk, and A. A. Clerk for fruitful discussions and comments, and for careful reading of the manuscript. **Funding:** K.C.M, J.P.B., C.P.A, A.B., A.L.C, G.W., and D.D.A. were supported by DARPA D18AC00015KK1932, AFOSR FA9550-19-1-0358, ONR N00014-17-1-3026, and the UChicago MRSEC NSF DMR-1420709. H.A. and T.O. were supported by JSPS KAKENHI 18H03770 and 20H00355. This work made use of the Pritzker Nanofabrication Facility part of the Pritzker School of Molecular Engineering at the University of Chicago, which receives support from Soft and Hybrid Nanotechnology Experimental (SHyNE) Resource (NSF ECCS-1542205), a node of the National Science Foundation's National Nanotechnology Coordinated Infrastructure. **Author contributions:** K.C.M. conceived and designed the experiment, and carried out theoretical calculations. K.C.M. and J.P.B. performed the experiments, with the assistance of G.W. C.P.A. annealed and fabricated the sample. K.C.M., J.P.B., A.B., and A.L.C. developed the confocal microscope setup. H.A. and T.O. performed electron irradiation of the SiC samples. D.D.A. advised on all fronts. All authors contributed to manuscript revision and preparation. **Competing interests:** The authors declare no competing financial interests. **Data and materials availability:** The datasets generated during and/or analysed during the current study are available from the corresponding author on reasonable request.

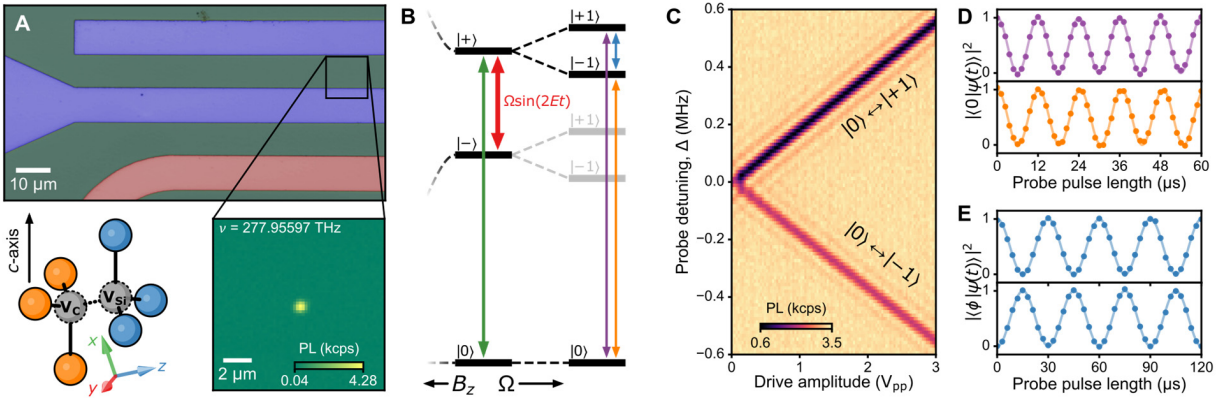

**Fig. 1. Driven *kh* divacancy spin system in 4H-SiC.** **(A)** False-color optical microscope image of the 4H-SiC sample showing an on-chip capacitor (blue) and wire (red) for microwave manipulation of the spin. Inset: Single *kh* divacancy under resonant optical excitation. Lower left: Lattice diagram of a *kh* divacancy and nearest-neighbor carbon atoms (blue) and silicon atoms (orange). Cartesian axes of the *kh* divacancy shown as colored vectors. **(B)** Energy diagram of the *kh* divacancy ground-state spin. At **B** = 0, clock transitions form between the three spin-1 levels. Spin manipulation in this basis consists of driving $|0\rangle \leftrightarrow |+\rangle$ (green) and $|+\rangle \leftrightarrow |-\rangle$ (red). Sufficiently strong driving of $|+\rangle \leftrightarrow |-\rangle$ induces Autler-Townes splitting, forming a hybridized spin-photon system, with an accessible upper branch and an inaccessible lower branch (gray). In this basis, spin manipulation consists of magnetically driving $|0\rangle \leftrightarrow |\pm 1\rangle$ (purple, orange), or electrically driving $|+1\rangle \leftrightarrow |-1\rangle$ (blue). **(C)** Pulsed optically detected magnetic resonance resolving Autler-Townes splitting of the ground-state spin. Δ is the probe frequency detuning from the $|0\rangle \leftrightarrow |+\rangle$ resonance frequency. **(D)** Magnetically driven Rabi oscillations using the $|0\rangle \leftrightarrow |+1\rangle$ (upper) and $|0\rangle \leftrightarrow |-1\rangle$ (lower) transitions corresponding to the purple and orange arrows in (B). **(E)** Electrically driven Rabi oscillations using the transition $|+1\rangle \leftrightarrow |-1\rangle$ highlighted by the blue arrow in (B) after reading out the population of $|+1\rangle$ (φ = +1) (upper) and $|-1\rangle$ (φ = −1) (lower).

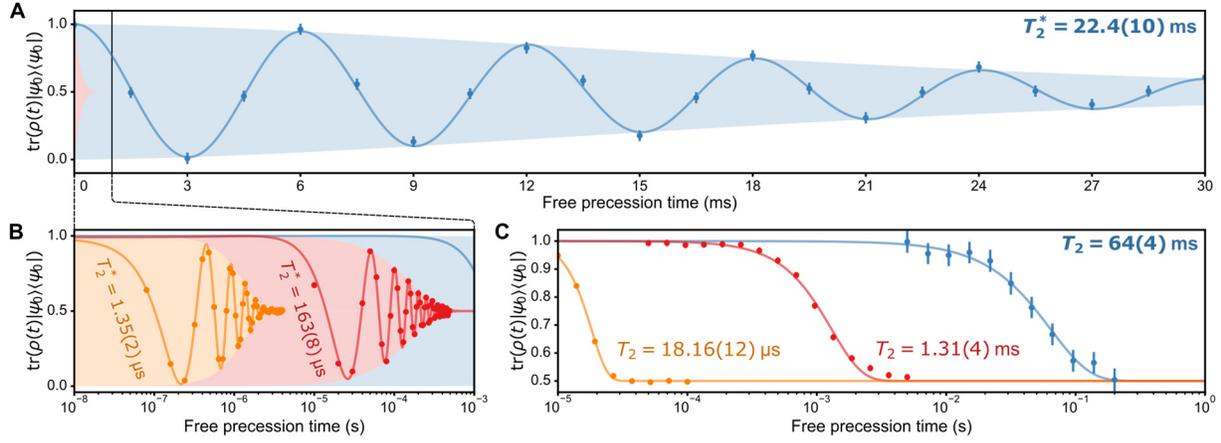

**Fig. 2. Spin coherence in the DPS. (A)** Ramsey free precession of $|\psi_0\rangle = \frac{1}{\sqrt{2}}(|+1\rangle + |-1\rangle)$ in the DPS at **B** = 0 mT (blue). A microwave detuning of +166.6 Hz is added to increase visibility of the decay envelope. **(B)** Ramsey free precession of $|\psi_0\rangle = \frac{1}{\sqrt{2}}(|0\rangle + |+\rangle)$ (red) and $|\psi_0\rangle = \frac{1}{\sqrt{2}}(|0\rangle + |+1_z\rangle)$ (orange) prepared outside of the DPS at **B** = 0 mT and $B_z$ = 1.2 mT, respectively. **(C)** Hahn echo free precession of the spin when prepared in the DPS at **B** = 0 mT (blue, $|\psi_0\rangle = \frac{1}{\sqrt{2}}(|+1\rangle + |-1\rangle)$), and outside of the DPS at **B** = 0 mT (red, $|\psi_0\rangle = \frac{1}{\sqrt{2}}(|0\rangle + |+\rangle)$) and at $B_z$ = 1.2 mT (orange, $|\psi_0\rangle = \frac{1}{\sqrt{2}}(|0\rangle + |+1_z\rangle)$). Error bars represent 1 SD.

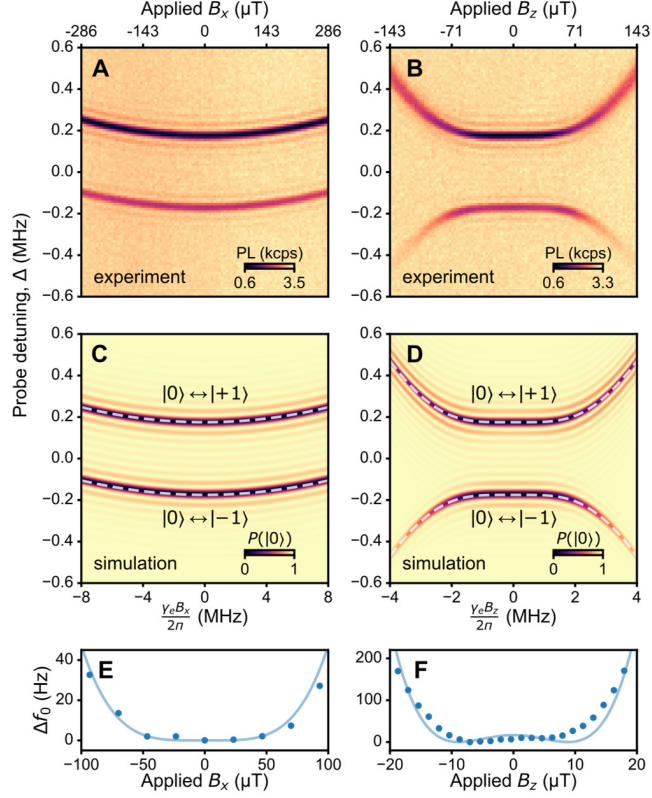

**Fig. 3. Energy dispersion in the DPS. (A, B)** Measured spectrum of the $|0\rangle \leftrightarrow |\pm 1\rangle$ transitions over a range of applied *x*-axis (A) and *z*-axis (B) magnetic fields. The contrast recovery procedure (see Supplementary Materials) produces unequal photoluminescence contrast between the two resonances. The probe detuning, $\Delta$, is referenced to the $|0\rangle \leftrightarrow |+\rangle$ resonance frequency. In (B), high field induces inhomogeneous broadening, as the spin is no longer fully encoded in the DPS. **(C, D)** Simulated spin resonance spectrum of the driven $|\pm 1\rangle$ states corresponding to (A) and (B), respectively. Dashed white lines indicate transition energy spectra derived from Floquet analysis. **(E, F)** Energy difference $\Delta f_0$ between the $|+1\rangle$ and $|-1\rangle$ states as a function of applied $B_x$ (E) and applied $B_z$ (F), as measured using Ramsey interferometry of $|\psi_0\rangle = \frac{1}{\sqrt{2}}(|+1\rangle + |-1\rangle)$. Solid lines are the energy differences derived from Floquet analysis with no free parameters. Error bars are smaller than the symbol size.

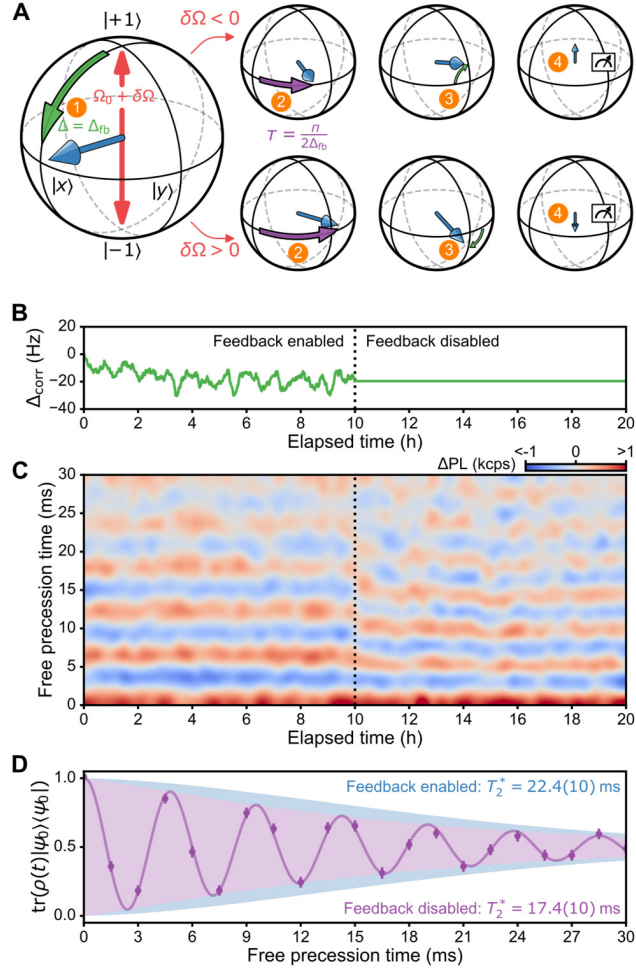

**Fig. 4. Active spin resonance feedback. (A)** Bloch spheres illustrating the spin resonance feedback protocol. See Methods for step-by-step details. **(B)** Applied corrective detuning to the frequency of the drive resonant with $|+1\rangle \leftrightarrow |-1\rangle$. During hours 0-10, active feedback is applied to the spin and the corrective detuning is derived from a proportional factor of the measured error signal. **(C)** Stabilization of Ramsey free precession of $|\psi_0\rangle = \frac{1}{\sqrt{2}}(|+1\rangle + |-1\rangle)$ with active spin resonance feedback. While active feedback is enabled, the observed Ramsey fringes are stable at the chosen detuning of +166.6 Hz. Once disabled, slow drifts in the dressing drive Rabi frequency shift the effective detuning of the microwave pulse used to prepare the superposition. Interpolation applied to emphasize fringe locations. **(D)** The Ramsey free precession decay envelope when integrating the individual Ramsey free precession iterations in (C), demonstrating the increased inhomogeneity without the feedback process. Error bars represent 1 SD.